# Strain-Enhanced Hydrogen Evolution, Electrical, Optical, and Thermoelectric Properties of the Multifunctional 2D $CrSi_2N_4$ Monolayer

Rao Uzair Ahmad[1,a] , Fahd Sikandar Khan[2,b], Nasir Javed[1,c]

[1] Ghulam Ishaq Khan Institute of Engineering Sciences and Technology, Tarbela Road, Topi, 23460, Pakistan

[2] National University of Science and Technology, Islamabad, Pakistan

Email: [a] rao.uzair@giki.edu.pk, [b]fahd.khan@seecs.edu.pk, , [c]nasir.javed@giki.edu.pk

## Abstract

First-principles density functional theory (DFT) is employed to evaluate the structural, electronic, optical, thermoelectric, and electrocatalytic properties of monolayer $CrSi_2N_4$. Its symmetric N-Si-N-Cr-N-Si-N septuple-layer structure exhibits dynamic, thermal (300 K), and mechanical stability, supported by a -8.76 eV/atom cohesive energy. PBE and HSE06 functionals reveal an indirect bandgap of 0.58 eV and 2.16 eV, respectively, driven by localized Cr-3d and N-2p states. The monolayer features 15.57 static dielectric constant and maximum absorption coefficients of $0.9\times10^{6}$ $cm^{-1}$ (visible) and $1.4\times10^{6}$ $cm^{-1}$ (deep-UV). Semiclassical Boltzmann calculations predict an outstanding room-temperature n-type thermoelectric power factor of 3.5 x $mW/mK^{2}$. For hydrogen evolution (HER), the basal plane yields a baseline hydrogen adsorption free energy ($\Delta G_H$) of 1.05 eV at the N-site. Applying +5% expansive biaxial strain improves HER kinetics, reducing $\Delta G_H$ to 0.46 eV. Thus, $CrSi_2N_4$ is a resilient, tuneable candidate for waste-heat recovery, photodetectors, and sustainable electrocatalysis.

## Keywords

2D CrSi2N4, DFT, Hydrogen Evolution Eeaction, Strain Engineering, Thermoelectric Properties.

## Introduction

The quest for the sustainable energy solutions has intensified in response to escalating global energy demands and environmental crises linked to fossil fuel consumption. Among renewable energy carriers, hydrogen stands out as a zero-emission fuel with high energy density (142 MJ/kg), positioning itself as a cornerstone for decarbonising industries, transportation, and power generation [1]. However, scalable hydrogen production via water electrolysis remains constrained by the reliance on platinum-group metals (PGMs) as catalysts for the hydrogen evolution reaction (HER), which are prohibitively expensive and scarce [2]. This challenge has spurred extensive research into earth-abundant alternatives, with two-dimensional (2D) materials emerging as promising candidates due to their unique electronic properties, high surface-to-volume ratios, and tuneable catalytic activity [3].

The exploration of two-dimensional (2D) materials has fundamentally transformed the landscape of nanoscale device engineering and sustainable energy conversion [4]. While typical 2D materials such as graphene, transition metal dichalcogenides (TMDs), and MXenes have demonstrated exceptional physical and chemical properties, they frequently encounter practical limitations in large-scale applications. For example, the catalytic activity of $MoS_2$ is largely confined to its edge sites while the vast basal plane remains inert [5], whereas high-conductivity MXenes often suffer from surface oxidation and structural degradation under ambient or operational conditions [6].

A significant breakthrough addressing these issues occurred in 2020 with the successful chemical vapor deposition (CVD) synthesis of continuous monolayer $MoSi_2N_4$ [7]. This synthesis was achieved by introducing elemental silicon during the growth of non-layered molybdenum nitride. This effectively passivated the surface and forced its crystallization into a highly stable 2D van der Waals structure. This milestone established a completely new structural paradigm of expansive $MA_2Z_4$ family of materials. Where M represents an early transition metal, A represents a group-IV element like Si or Ge, and Z represents a pnictogen such as N, P, or As. The defining feature of $MA_2Z_4$ monolayers is their complex septuple-layer morphology. This structure consists of a central transition metal-pnictogen core ($M-Z_2$) seamlessly intercalated between two protective group-IV-pnictogen outer sheaths ($A-Z$). This unique architecture serves a profound physical purpose, giving it exceptional mechanical resilience, high isotropic Young's moduli, and intrinsic resistance to ambient scattering and spontaneous surface oxidation. Moreover, their tuneable electronic properties, layer structures and high stability contribute to the application in optoelectronics[8,9], thermal management systems[10], spintronics[9,11] and electronics[7].

Since the emergence of $MoSi_2N_4$ [7], numerous first-principle studies have explored structural, optical and electronic properties of 2D $MA_2Z_4$ materials [8,12,13]. Various approaches have been proposed to engineer these properties, with external electric fields and mechanical strain being among the most commonly examined strategies [14–16]. Beyond pristine configurations, the introduction of defects and doping into $MoSi_2N_4$ has also gained considerable computational attention [17–19] Practically, these modifications are often aimed at catalytic applications, with MoSi2N4 demonstrating strong potential in hydrogen [17,20,21] and oxygen evolution [22], oxygen reduction [22] and the electrocatalytic reduction of CO2 [19,23] and CO [18].

$CrSi_2N_4$ is promising to show excellent mechanical and thermal and catalytic stability. This is because the chromium-based nitrides, such as CrN and $Cr_2N$, are well-known for their hardness and oxidation resistance, but their catalytic performance has been limited by poor electrical conductivity [23]. The $CrSi_2N_4$ consists of hexagonal structure comprising a $CrN_2$ layer sandwiched between Si-N bilayers. This confers mechanical resilience and resistance to surface passivation, critical traits for electrocatalysts operating in acidic or alkaline media [24].Preliminary studies on analogous structures, such as $MoSi_2N_2P_2$, have demonstrated promising HER activity with Gibbs free energies ($\Delta G_H$) close to the thermoneutral ideal ($\Delta G_H \approx 0\ eV$) [25], raising the possibility that $CrSi_2N_4$ could exhibit similar or superior performance. Despite these insights, investigations into the HER mechanism of $CrSi_2N_4$ remain absent. Key unanswered questions include the role of surface terminations (e.g., N-vacancies, functional groups) in modulating active sites, the impact of strain or doping on charge transfer kinetics, and the material's stability under electrochemical conditions. Addressing these gaps requires a multidisciplinary approach combining density functional theory (DFT) simulations, electronic structure analysis, and experimental validation.

This study employs DFT calculations to evaluate the HER potential of 2D $CrSi_2N_4$. Through comprehensive first principles calculations, we demonstrate that $CrSi_2N_4$ is dynamically, thermally and mechanically stable, supported by an exceptionally high cohesive energy of -8.76 eV/atom.

Utilizing the HSE06 hybrid functional, we confirm that the material is an indirect bandgap semiconductor (2.16 eV) exhibiting excellent optical absorption coefficients reaching $1.4 \times 10^{6}\ cm^{-1}$ in the deep-UV region. Finally, the $\Delta G_{H}$ calculations were performed to estimate its HER performance. Furthermore, we benchmark $CrSi_2N_4$ against established catalysts like Pt(111) and $MoSi_2N_4$ to contextualize its performance. We got 58.16% enhancement in HER performance for $CrSi_2N_4$ as compared to $MoSi_2N_4$. To further reduce the HER overpotential we applied the biaxial strain to the monolayer.

**Computational Method:**

First-principles DFT calculations were performed using Quantum Espresso V.7.3 [26] (a plane-wave basis set code) using optimized norm-conserving Vanderbilt pseudopotentials (ONCVPSP) [27]. The kinetic energy cutoff to expand plane waves was set to 80 Ry, and the Γ-centred k-point grid to 11 × 11 × 1 for the first Brillouin zone was employed. For relaxations of the structure with the energy convergence of $>10^{-6}$ Ry and the force convergence was set to $10^{-3}$ Ry/Bohr. Structural optimizations were carried out by employing the generalized gradient approximation (GGA) with the PBE flavour [28] for electron exchange-correlation functionals. Furthermore, as the structure was two-dimensional, a wide vacuum spacing (~20 Å) was included along the z-axis to depict a similar experimentally available system to minimize possible interactions between the periodic images of the structure. To compensate for the underestimation of GGA-PBE in predicting band gap energies, we used the Heyd-Scuderia-Ernzerhof (HSE06) hybrid functional [29]. Electronic band structures with the HSE06 functional were obtained via Wannierization with Wannier90 V.3.1.0 [30–32]. To set up this process, Quantum ESPRESSO's open_grid.x executable was first run to construct the Kohn-Sham state data. Subsequently, the pw2wannier.x code was used to interface these wavefunctions with Wannier90. Finally, the band structures were calculated by employing the Wannier90 code. To evaluate the dynamic stability of the monolayer $CrSi_2N_4$, the phonon spectrum was calculated using the density functional perturbation theory method implemented in the PHONOPY package [33]. Cohesive energy was calculated to predict the stability of the $CrSi_2N_4$, and the room temperature thermal stability was investigated by performing the ab-initio molecular dynamics calculations using the Anderson thermostat. The thermoelectric properties were calculated using the Boltzmann transport equation in the relaxation time approximation implemented by the BoltzWann code [34].

**Results and Discussion**

*Structural Properties and Stabilities:*

The optimized crystal structure of a $CrSi_2N_4$ unit cell with the stacking order of N-Si-N-Cr-N-Si-N atomic layers is shown in Figure 1. It has a hexagonal structure with space group $P\bar{6}m2$. The computed optimized lattice constant is 2.8440 Å which is less than that of $Si_3N_4$ (~3.1 Å). This contraction could be due to the strong bonding between Cr and N and is comparable to a previously reported structure [7]. The chromium atoms occupy high-symmetry sites with Cr–N bond length of $\sim$1.98 Å. Silicon atoms are positioned in a honeycomb arrangement with Si–N average bond distance of $\sim$1.72 Å, creating a puckered sublattice that modulates the in-plane strain. The detailed structural parameters can be seen in Table 1.

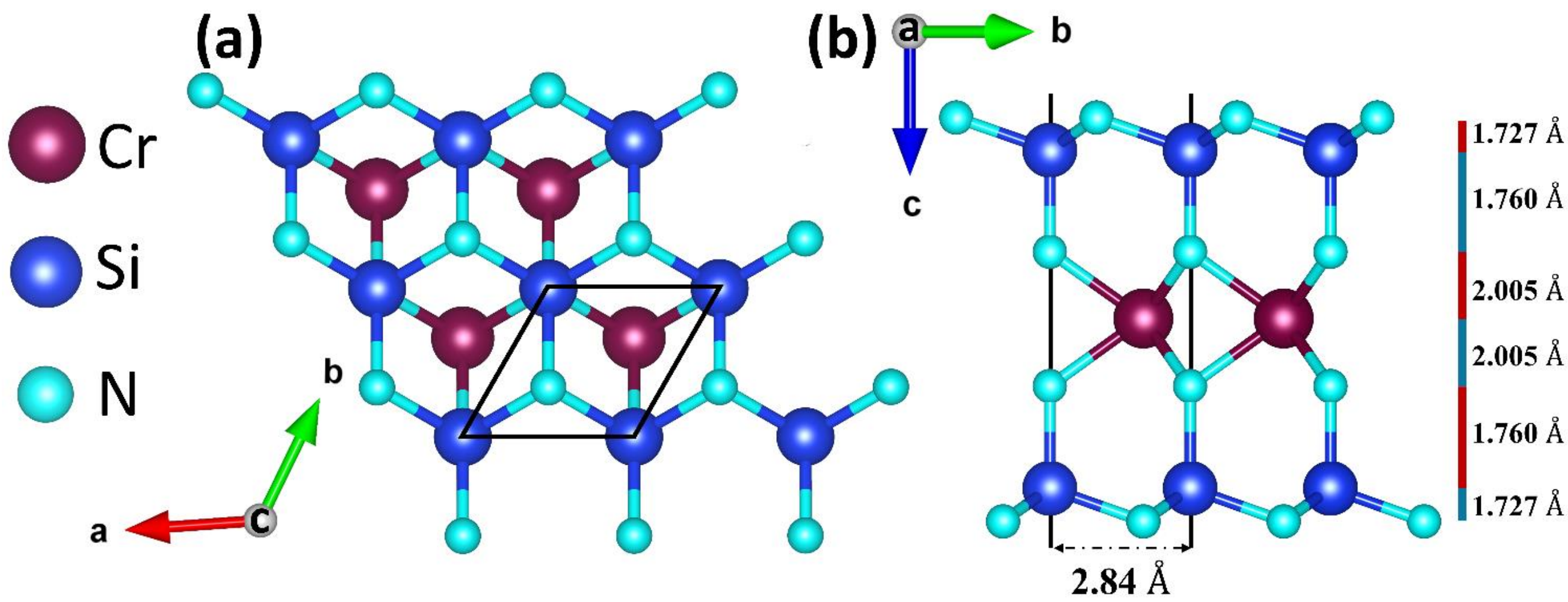

**Figure 1. (a) Top and (b) side views of the optimized $CrSi_2N_4$ septuple-layer monolayer (N-Si-N-Cr-N-Si-N). High-symmetry bond lengths are indicated for the Cr-N (1.98 Å) and Si-N (1.71–1.74 Å) interfaces.**

**Table 1. Calculated structural parameters of the $CrSi_2N_4$ monolayer, including lattice constants, bond lengths, and internal angles.**

| **Parameter / Property** | **Calculated Value** |
|---|---|
| Crystal System | Hexagonal |
| Space Group | $P\bar{6}m2$ |
| Stacking Sequence | N-Si-N-Cr-N-Si-N |
| Lattice Constant (a) | 2.844 Å |
| Bond Length (Cr-N) | 1.98 Å |
| Outer Layer Bond Length (Si-N) | 1.71 Å |
| Inner Layer Bond Length (Si-N) | 1.74 Å |
| Internal Angle (N2-Cr-N1) | 69.75° |
| Internal Angle (Cr-N2-Si2) | 72.50° |
| Internal Angle (N4-Si-N4) | 110.85° |
| Internal Angle (N1-Si-N4) | 108.05° |

It is well known that structural stability is essential for practical applications. To confirm this, we used three approaches: cohesive energy, phonon dispersion and the ab-initio molecular dynamics (AIMD) evolution. Cohesive energy was calculated using the relation.

$$E_{cohesive} = \frac{E_{CrSi2N4} - N_{Cr}E_{Cr} - N_{Si}E_{Si} - N_N E_N}{N_{Cr} + N_{Si} + N_N}$$

Where $E_{Cr}$, $E_N$, *and* $E_{Si}$ are the energies of Cr, N, and Si atoms, respectively. $E_{CrSi2N4}$ is the total energy of the $CrSi_2N_4$ . Also, $N_{Cr}$, $N_N$, *and* $N_{Si}$ represent the number of atoms of Cr, N, and Si in the unit cell, which are as follows:$N_{Cr} = 1$, $N_{Si} = 2$, $N_N = 4$

The calculated cohesive energy of $CrSi_2N_4$ is $-8.76$ eV/atom, which is comparable to the experimentally measured cohesive energies of $MoSi_2P_4$ ($-6.21$ eV) and $MoSi_2N_4$ ($-8.55$ eV) [7]. This confirms the stability of the $CrSi_2N_4$ structure, the high cohesive energy of the studied $CrSi_2N_4$ structure, which is comparable to the cohesive energy of $MoSi_2N_4$, is the indication of the structural stability of the $CrSi_2N_4$ structure. Moreover, this cohesive energy is orders of magnitude larger than the ambient thermal energy at standard conditions ($k_B\, T \approx 26\, meV$). This confirms that the structure is highly stable and resistant to thermal degradation at room temperature.

A favourable cohesive energy does not completely guarantee an absence of the structural phase transition, lattice collapse or vibrational anomalies. Therefore, dynamic stability must be assessed by the phonon dispersion spectra. To investigate the dynamic stability of the material, its phonon dispersion band structure was calculated. $1 \times 1 \times 1$ unit cell was used for this purpose. The phonon band structure in Figure 2a shows the 21 phonon branches as there is 7 atoms in the unit cell, 3 acoustic and 18 optical branches. Theoretically, the existence of soft phonon modes which are mathematically indicated by imaginary frequencies ($\omega^2 < 0$) across high-symmetry paths of Brillouin zone signals an unphysical structural stability where atoms would spontaneously distort away from the equilibrium position without encountering a restorative force. Our spectra show no imaginary frequencies at any point even with the choosing the small supercell. Therefore, this suggests that $CrSi_2N_4$ is dynamically stable for different applications.

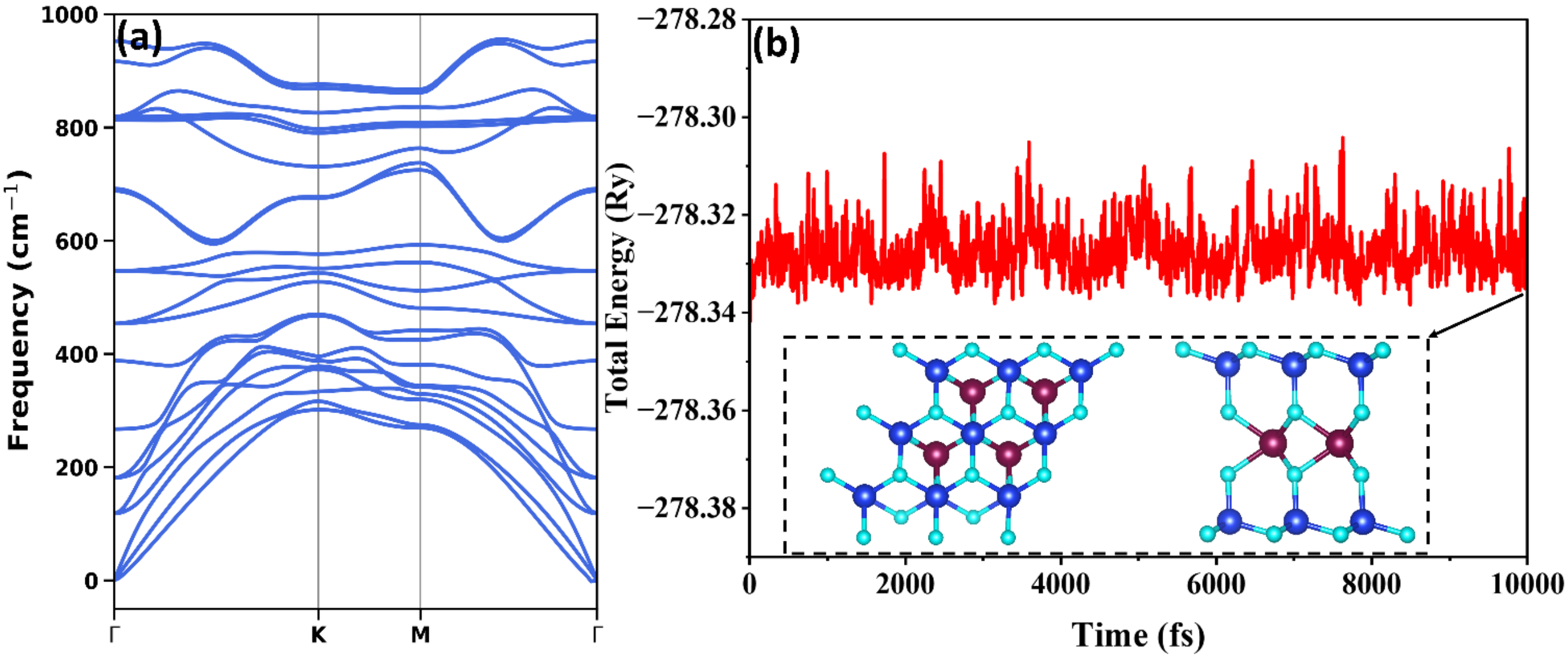

**Figure 2. Dynamic and thermal stability of the $CrSi_2N_4$ monolayer. (a) Phonon dispersion spectra displaying the absence of imaginary frequencies. (b) Ab-initio molecular dynamics (AIMD) evolution of total energy over 10 ps at 300 K.**

The AIMD evolution of the structure was calculated at 300K over 10 ps to evaluate the thermal stability at room temperature. The simulations were carried out in the canonical (NVT) ensemble using the Andersen thermostat to regulate the system temperature at 300 K through stochastic velocity rescaling. As illustrated in Figure 2b, the total energy exhibited no considerable fluctuations over the complete duration of the simulation. Because the atoms largely retained their initial relaxed configurations without undergoing severe distortions, we can confidently conclude that the material is thermally stable under ambient room-temperature conditions.

**Table 2. The calculated mechanical properties of CrSi2N4 and their comparison with graphene.**

| | $CrSi_2N_4$ | Graphene |
|---|---|---|
| **2D Young's modulus ($Y_{2D}$) [N/m]** | 468.75 | 340 [35] |
| **2D shear modulus ($G_{2D}$) [N/m]** | 182.13 | 148.7 [36] |
| **2D Poisson's ratio ($\nu$)** | 0.2869 | 0.185 [36] |

We further evaluated the mechanical stability of the $CrSi_2N_4$ structure. The elastic constants ($C_{lm}$) were calculated for this purpose. The Born's criteria were also tested, which is $C_{11} > 0$ and $C_{11}^2 - C_{12}^2 > 0$ [36]. The calculated $C_{11}$ and $C_{12}$ are 510.79 N/m and 146.52 N/m respectively. The 2D Young's modulus ($Y_{2D}$), Poisson ratio ($\nu$) and the shear modulus ($G_{2D}$) can be calculated according to literature as [37],

$$Y_{2D} = \frac{C_{11}^2 - C_{12}^2}{C_{11}}, G_{2D} = C_{66}, \nu = \frac{C_{12}}{C_{11}}$$

The calculated mechanical properties reveal that the $CrSi_2N_4$ monolayer possesses excellent in-plane stiffness, significantly outperforming graphene, which is widely recognized as one of the most mechanically robust 2D materials. As shown in Table 2, the 2D Young's modulus of $CrSi_2N_4$ (468.75 N/m) is approximately 38% higher than the literature value for graphene. This mechanical rigidity is because the monoatomic thickness of graphene's purely planar $sp^2$ carbon lattice. On the other hand, $CrSi_2N_4$ features a thick, septuple-layer with cross-linked covalent bonding network effectively which distributes applied longitudinal stresses across structure, leading to an exceptionally high resistance to tensile strain. Similarly, the 2D shear modulus of $CrSi_2N_4$ (182.13 N/m) notably exceeds that of graphene (148.7 N/m). Th shear's modulus decides the material resistance to permanently deform under the applied shear stress. The calculated higher values shows that the resistance of $CrSi_2N_4$ against shear deformation is higher than the Graphene.

Furthermore, the calculated Poisson's ratio for $CrSi_2N_4$ ($\nu = 0.2869$) falls within the expected range for transition metal-based 2D compounds [37], though it is notably higher than the Poisson's ratio of graphene (0.185). This higher $\nu$ value signifies a more pronounced transverse contraction when the material is subjected to uniaxial longitudinal stretching. This mechanical response is a direct consequence of the complex internal degrees of freedom within the $CrSi_2N_4$ unit cell [8]. While the material is extremely stiff against total area expansion, the individual atomic sub-layers have the capacity to undergo slight internal relaxations and out-of-plane buckling under strain [38], facilitating a greater transverse dimensional change than a rigid, single-atom layer.

*Electronic Band structure:*

Understanding the electronic band structure of $CrSi_2N_4$ is critical for assessing its fundamental capability to participate in the charge transfer processes, particularly those required for the optoelectronics, thermoelectric and surface electrocatalysis. The calculated electronic properties are shown in Figure 3. Initial bandgap estimation was performed using the standard GGA-PBE functional, which predicts a narrow indirect bandgap of 0.58 eV for the monolayer with no net spin. The valence band maximum (VBM) is located at the Gamma (Γ) high-symmetry point, while the conduction band minimum (CBM) is located at the K point.

To visualize the contribution of each atomic layer to the band structure, the band structure was projected on to each atom individually. Moreover, the atom-projected density of states (DOS) was calculated. The atomic layer projected band structure shows that the states close to the fermi level are dominated by the Cr and N atoms (mostly Cr atom). In the $MA_2Z_4$ family, the central transition metal's d-orbitals are subjected to an intense, highly symmetric crystal field generated by the surrounding nitrogen polyhedra. For $CrSi_2N_4$, the electronic states immediately flanking the Fermi energy ($E_F$) specifically the CBM and VBM are almost dominated by the highly localized Cr 3d states. These Cr 3d orbitals do not exist in pure isolation, but they undergo, energy-lowering covalent hybridization with the 2p states of the directly coordinating inner nitrogen atoms. This 3d-2p orbital mixing dictates the width, contour, and energetic dispersion of the energy bands near the Fermi level.

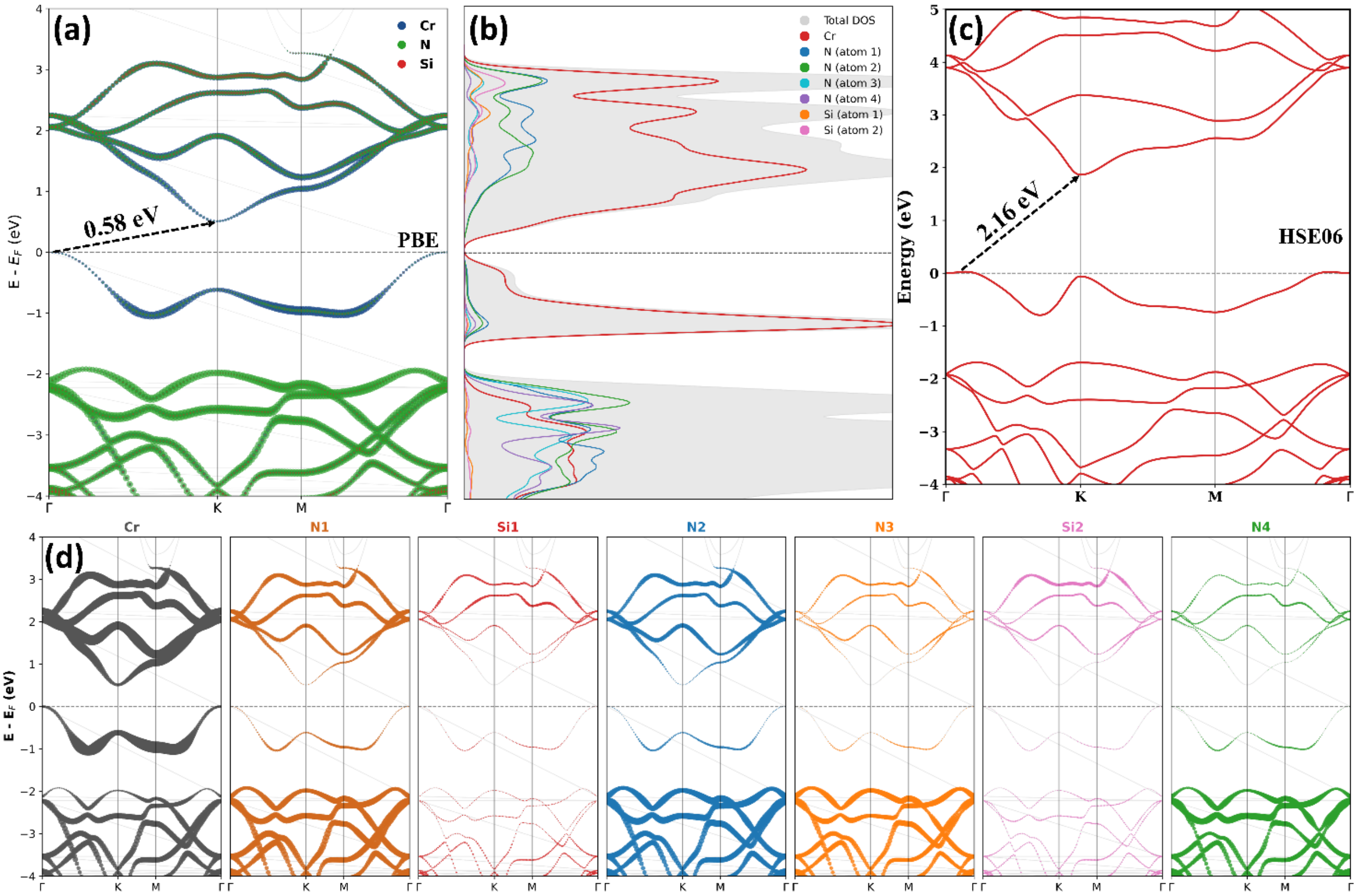

**Figure 3. Electronic properties of monolayer $CrSi_2N_4$. (a, b) Band structures and density of states calculated using GGA-PBE (c) Band structure calculated using HSE06 hybrid functionals, revealing an indirect bandgap of 2.16 eV. (d) Atom-projected density of states (PDOS) highlighting the dominant Cr-3d and N-2p orbital contributions near the Fermi level**

As the GGA-PBE functional is known to underestimate the electronic band gap, we also used the HSE06 hybrid functional to calculate the electronic band structure. The HSE06 calculation yields an increased band gap of 2.16 eV. Hong et al. [7] reported the band gap of a similar material, $MoSi_2N_4$ monolayer to be 1.94 eV. They also reported band gap values of 1.744 eV and 2.297 eV for the $MoSi_2N_4$ monolayer using the PBE and HSE06 hybrid functionals, respectively. It is worth noting that recent computational studies using the GW approximation have reported larger band gaps of 2.82 eV for the monolayer, 2.67 eV for the bilayer, and 2.41 eV for bulk $MoSi_2N_4$ [39]. Since the GW approximation typically overestimates the band gap, the experimental value is expected to lie between the PBE and HSE06 predictions.

*Thermoelectric Properties:*

To evaluate the potential of the 2D $CrSi_2N_4$ monolayer for energy harvesting applications, its macroscopic thermoelectric transport properties were computed along xx-direction using semiclassical Boltzmann transport theory within the constant relaxation time approximation (CRTA). The transport coefficients such as the Seebeck coefficient (S), electrical conductivity ($\sigma$), electronic thermal conductivity ($\kappa_e$), and power factor (PF) were evaluated as a function of the chemical potential ($\mu - E_F$) across a temperature range of 300 K to 700 K.

Figure 4a illustrates the Seebeck coefficient of the $CrSi_2N_4$ monolayer. The profile exhibits a classic bipolar semiconductor signature, with the Seebeck coefficient crossing zero exactly at the Fermi level due to the perfect cancellation of electron and hole thermopower. At 300 K, the material demonstrates exceptionally high thermopower, reaching maximum absolute values of 1108 $\mu$V/K for p-type doping (at -0.05 eV) and -1229 $\mu$V/K for n-type doping (at +0.05 eV). As the temperature increases from 300 K to 700 K, the maximum absolute Seebeck values systematically decrease and the peaks broaden. This suppression at higher temperatures is a well-documented phenomenon driven by thermal smearing, as $k_{BT}$ increases, the Fermi-Dirac distribution widens, exciting minority carriers across the bandgap, which fundamentally counteracts the primary majority-carrier voltage.

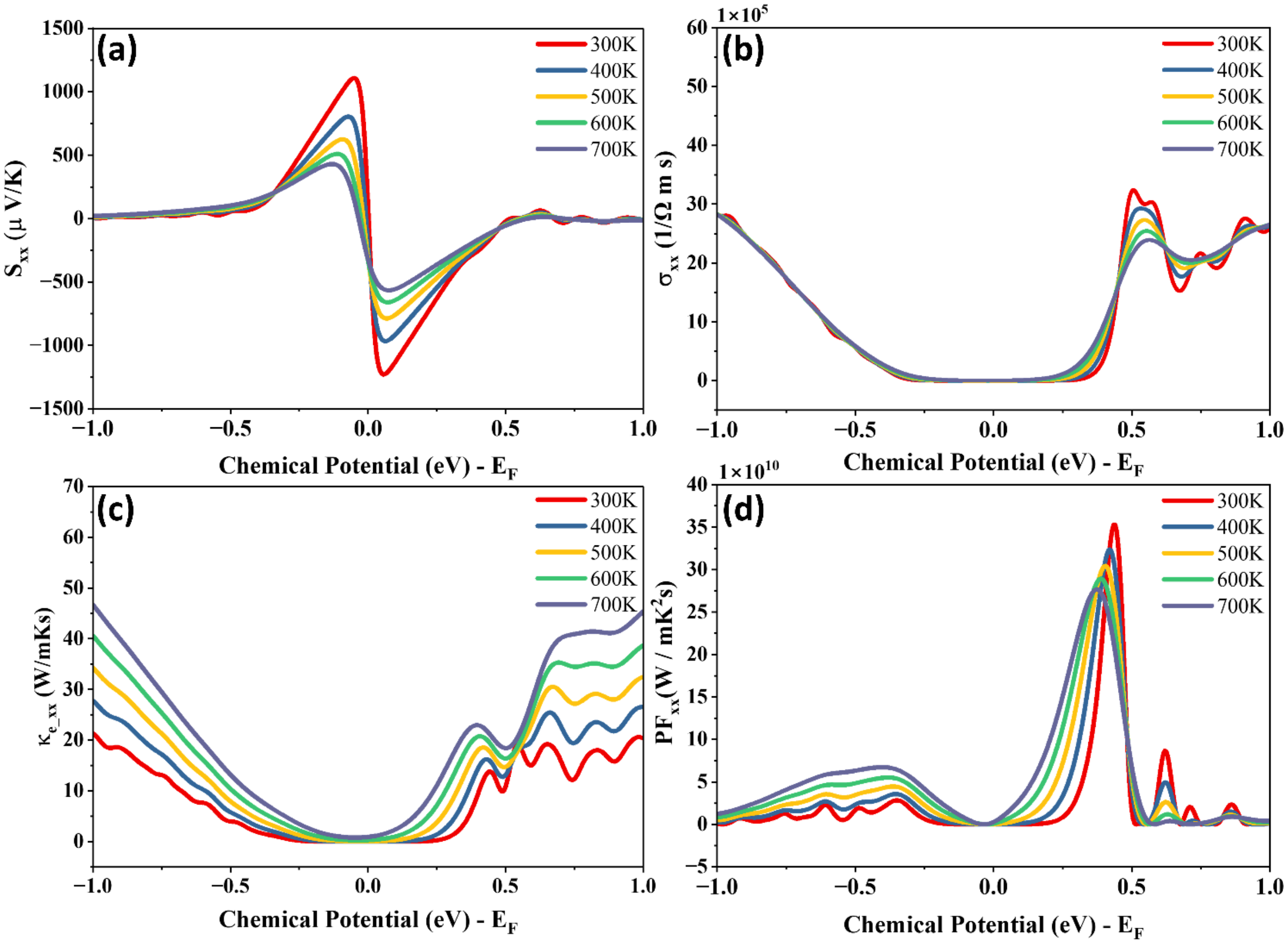

**Figure 4. Calculated thermoelectric transport coefficients as a function of chemical potential ($\mu - E_F$) from 300 K to 700 K: (a) Seebeck coefficient, (b) electrical conductivity, (c) electronic thermal conductivity, and (d) power factor.**

The electrical conductivity with respect to the relaxation time ($\sigma/\tau$) is presented in Figure 4b. In the intrinsic region spanning roughly -0.3 eV to +0.3 eV, the electrical conductivity drops strictly to zero, confirming the presence of a robust electronic bandgap where no charge carriers are available for transport. As the chemical potential is shifted deeper into the valence band (negative $\mu$) or conduction band (positive $\mu$), $\sigma/\tau$ experiences a rapid, monotonic increase due to the population of delocalized electronic states. The density of states (DOS) near the band edges heavily dictates this slope. The slightly steeper onset on the conduction-band side suggests a higher group velocity for electrons compared to that of holes.

Figure 4c displays the electronic thermal conductivity ($\kappa_e/\tau$). The structural profile of the $\kappa_e$ curves is nearly identical to that of the electrical conductivity, a direct manifestation of the Wiedemann-Franz law:

$$\kappa_e = L\,\sigma\,T$$

where $L$ is the Lorenz number. Because the heat transported by charge carriers is directly proportional to both the electrical conductivity and the absolute temperature, $\kappa_e$ scales upward significantly as the system is heated to 700 K. The power factor ($PF = S^2\sigma$), which serves as the primary metric for the electrical performance of a thermoelectric device, is plotted in Figure 4d. The optimization of the power factor requires a delicate balance, as heavy doping increases $\sigma$ but inherently destroys the Seebeck coefficient, $S$. The calculated $PF/\tau$ reveals a striking electron-hole asymmetry in the transport behavior of $CrSi_2N_4$. For p-type doping, the power factor reaches a moderate local maximum of approximately $3 \times 10^{10} Wm^{-1}K^{-2}s^{-1}$ near $\mu - E_F = -0.3\ eV$. In contrast, optimal n-type doping yields a colossal peak of $35 \times 10^{10} Wm^{-1} K^{-2} s^{-1}$ at $\mu - E_F \approx +0.43\ eV$ at 300 K.

To ground these computational values in experimental reality, we apply a standard empirical relaxation time for 2D semiconductors of $\tau = 10\ fs$ [40]. Under this approximation, the theoretical n-type peak translates to an absolute power factor of 3.5 $mW\ m^{-1}\ K^{-2}$. This magnitude is highly competitive with state-of-the-art commercial thermoelectric materials such as $Bi_2Te_3$, clearly demonstrating that n-doped $CrSi_2N_4$ is a promising candidate for next-generation, low-dimensional waste heat recovery devices.

*Optical Properties:*

To evaluate the potential of the two-dimensional $CrSi_2N_4$ monolayer for optoelectronic and nanophotonic applications, the macroscopic optical properties were systematically investigated. The fundamental quantity dictating the linear optical response is the complex frequency-dependent dielectric tensor, $\varepsilon(\omega) = \varepsilon_1(\omega) + i\varepsilon_2(\omega)$. Because the calculations were performed using periodic boundary conditions within a 3D supercell, the raw dielectric output was artificially diluted by the vacuum layer introduced to isolate the monolayer. To extract the true optical response of the 2D sheet, a volumetric scaling correction was applied using:

$$\varepsilon^{2D}(\omega) = 1 + [\varepsilon^{3D}(\omega) - 1]\left(\frac{c}{t_{eff}}\right)$$

where $c$ is the out-of-plane lattice parameter and $t_{eff}$ is the effective geometric thickness of the septuple layer, including the van der Waals radii of the surface nitrogen atoms.

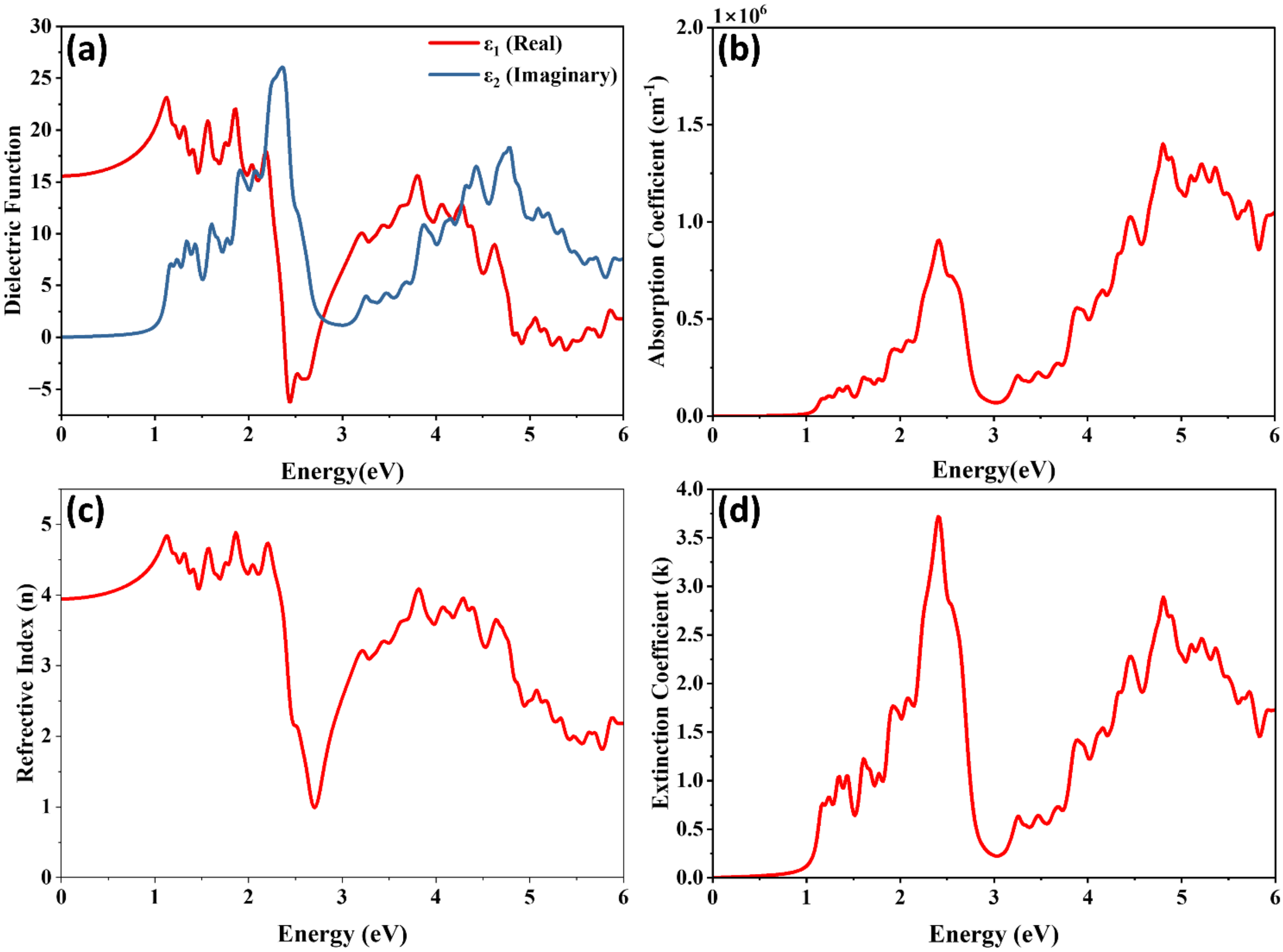

**Figure 5. The optical properties of the $CrSi_2N_4$ monolayer. (a) Real ($\epsilon_1$) and imaginary ($\epsilon_2$) parts of the dielectric function. (b) Optical absorption coefficient ($\alpha$). (c) Refractive index ($n$). (d) Extinction coefficient ($k$).**

The calculated real ($\varepsilon_1$) and imaginary ($\varepsilon_2$) parts of the in-plane dielectric function for the $CrSi_2N_4$ monolayer are presented in Figure 5a. The imaginary part, $\varepsilon_2(\omega)$ is directly proportional to the momentum matrix elements of transitions between occupied valence band and empty conduction band, which dictates the optical absorption behaviour. The optical onset occurs in the near infrared to visible boundary, followed by a prominent absorption peak centred at 2.35 eV. This intense resonant peak arises from direct interband transitions, likely dominated by the strongly correlated Cr-d and N-p orbitals near the Fermi level, a characteristic feature of transition-metal-based materials. The real part of the dielectric function, $\varepsilon_1(\omega)$, characterizes the polarization of the material and its ability to store electromagnetic energy. The static dielectric constant at the zero-frequency limit is observed to be notably high, with $\varepsilon_1(0) = 15.57$. By the principles of the Kramers-Kronig relations, the strong interband transition at 2.35 eV in $\varepsilon_2$ forces a sharp anomalous dispersion in $\varepsilon_1$. Consequently, $\varepsilon_1$ drops below zero between 2.37 eV and 2.76 eV. In this energy window, the monolayer behaves dynamically like a metal, perfectly reflecting incident photons. The energy crossing where $\varepsilon_1(\omega) = 0$ with a positive slope indicates the plasma frequency ($\omega_p$), where high-energy photons can once again propagate through the material. Further high-energy transitions are evident in the deep ultraviolet (UV) region (4.0-6.0 eV), causing secondary negative dips in $\varepsilon_1$.

**Table 3. Comparison of the calculated optical absorption coefficient ($\alpha$) of monolayer $CrSi_2N_4$ with other representative 2D and 3D optoelectronic materials.**

| | Dimensionality | $\alpha(cm^{-1})$ | Spectral Region | Reference |
|---|---|---|---|---|
| $CrSi_2N_4$ | 2D Monolayer | $9.0 \times 10^5$ | Visible (~2.4 eV) | This work |
| $CrSi_2N_4$ | 2D Monolayer | $> 1.4 \times 10^6$ | Deep UV (4.0 eV) | This work |
| $MoSi_2N_4$ | 2D Monolayer | $\sim 1.0 \times 10^6$ | Visible/UV | [7], [16] |
| $MoS_2$ | 2D Monolayer | $\sim 5.0 \times 10^5$ | Visible (~1.8 eV) | [41], [42] |
| $WS_2$ | 2D Monolayer | $\sim 6.0 \times 10^5$ | Visible (~2.0 eV) | [43] |
| $GaAs$ | Bulk (3D) | $\sim 1.0 \times 10^4$ | Near-IR/Visible | [44] |
| Silicon (Si) | Bulk (3D) | $\sim 10^3 - 10^4$ | Visible | [44,45] |

The practical light-harvesting efficiency of the monolayer is quantified by the optical absorption coefficient, $\alpha(\omega)$, shown in Figure 5b, calculated via the relation:

$$\alpha(\omega) = \frac{\sqrt{2}\omega}{c_{vac}} \sqrt{\sqrt{\varepsilon_1^2(\omega) + \varepsilon_2^2(\omega)} - \varepsilon_1(\omega)}$$

The absorption profile demonstrates that the $CrSi_2N_4$ monolayer possesses extraordinary light-harvesting capabilities. In the visible spectrum (2.4 eV), the absorption coefficient reaches an impressive maximum of $0.9 \times 10^6 cm^{-1}$ and in the deep UV region (at 4.8 eV), $\alpha$ reaches $1.4 \times 10^6\ cm^{-1}$. These values are roughly an order of magnitude higher than those of bulk silicon and are highly competitive with other emergent 2D transition metal dichalcogenides (TMDs). Table 3 shows the comparison of $CrSi_2N_4$ with other optoelectronic materials. Such ultra-high absorption coefficients suggest that even a single sub-nanometre layer of $CrSi_2N_4$ can interact robustly with incoming solar radiation, making it a highly promising candidate for ultrathin photovoltaic cells and UV photodetectors.

The complex refractive index, $N(\omega) = n(\omega) + ik(\omega)$, was further derived to evaluate the propagation of light through the lattice. Figure 5c and Figure 5d display the real refractive index ($n$) and the extinction coefficient ($k$), respectively. The static refractive index is determined to be $n(0) = 3.95$, which mathematically satisfies the relation $n(0) = \sqrt{\varepsilon_1(0)}$. As photon energy increases, $n(\omega)$ exhibits normal dispersion until the optical resonance at 2.21 eV, where it undergoes a drastic reduction, plunging to a minimum near $n = 1.00$. This deep minimum perfectly coincides with the primary maximum of the extinction coefficient ($k = 3.70$), reaffirming that light entering the material at this specific energy is attenuated and absorbed almost instantaneously. The strong optical anisotropy and robust visible to UV absorption cross-section indicate that strain engineering or surface functionalization of this structure could yield highly tuneable optoelectronic devices.

**Hydrogen evolution reaction (HER):**

Hydrogen gas ($H_2$) generated by the hydrogen evolution reaction (HER) follows the following reaction:

$$active\ site\ +\ H^+\ +\ e^-\ \rightarrow\ H^*$$

$$H^*\ +\ H^+\ +\ e^-\ \rightarrow\ H_2$$

Active site is the surface of the catalyst. H* represents the adsorbed hydrogen atom. The computational hydrogen electrode method [46] is used to calculate the hydrogen adsorption-desorption ($\Delta G_H$). We utilized the Gibbs free energy difference as a measure to evaluate the catalytic activity for the water-splitting HER. Therefore, $\Delta G_H$ of the structure is calculated by the following equation [47]:

$$\Delta G_H\ =\ \Delta E_H\ +\ \Delta E_{ZPE}\ -\ T\Delta S_H$$

Where $\Delta E_H$ is calculated computationally by quantum espresso, $\Delta E_{ZPE}$ is the zero point energy difference of hydrogen between the adsorbed state and the gas phase and $\Delta S_H$ is the entropy difference between the adsorbed and the gas phase. The values of the $\Delta E_{ZPE}$ and $T\Delta S_H$ was taken from the literature [48] and standard thermodynamic data [49] at $T = 298.15K$. $\Delta E_H$ is calculated by the relation:

$$\Delta E_H\ =\ E_{adsorbent\ +\ H}\ -\ E_{adsorbent}\ -\frac{1}{2}\ E_{H2}$$

Where $E_{adsorbent+H}$ is the energy of the adsorbent ($CrSi_2N_4$) with a hydrogen atom adsorbed on it, $E_{adsorbent}$ is the total energy of the adsorbent ($CrSi_2N_4$) and $E_{H2}$ is the total energy of the hydrogen molecule.

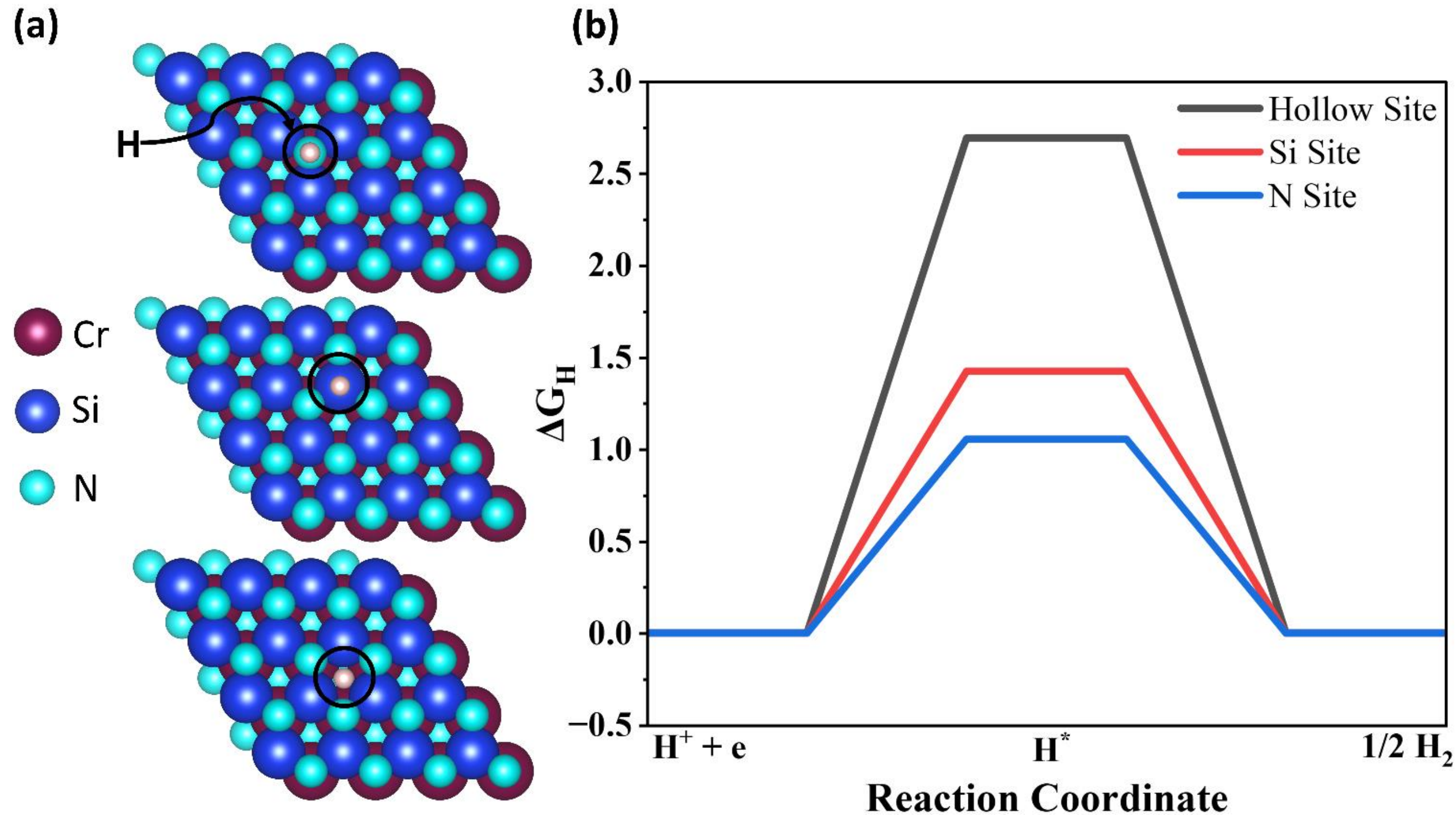

**Figure 6. Evaluation of the hydrogen evolution reaction (HER) performance. (a) Schematic of hydrogen adsorption at the N-site, Si-site, and hollow site. (b) Calculated Gibbs free energy ($\Delta G_H$) diagram for hydrogen adsorption at these sites.**

The hydrogen atom was adsorbed on the surface of the structure with two different sites and at the nitrogen vacancy Figure 6. A large enough supercell $(3 \times 3 \times 1)$ was used to avoid the interactions of hydrogen atoms with themselves by the periodic images. The HER performance of catalysts was analysed by determining the change in Gibbs free energy $(\Delta G_H)$ associated with hydrogen adsorption. A key observation is that maximum catalytic efficiency is attained when the magnitude of $\Delta G_H$ approaches negligible levels, signifying minimal energy difference between adsorbed hydrogen species and the reaction's participating phases. Excessively negative $\Delta G_H$ values correlate with overly stable hydrogen-catalyst bonds, complicating desorption, whereas overly positive values indicate insufficient binding strength for effective proton capture. These extremes collectively hinder reaction kinetics. The ideal standard for the $\Delta G_H$ is the Pt electrode, which has the value of $\approx 0eV$.

Figure 6 shows the plots for the Gibbs free energy for $CrSi_2N_4$ for HER on three different sites. The H atom was adsorbed on the N-site, the Si-site, and the hollow site. The adsorption of a hydrogen atom on the N-site gives the lower Gibbs free energy ($1.05\ eV$) as compared to the previously reported $MoSi_2N_4$ structure (~2.51 eV) [50]. $CrSi_2N_4$ also performs better than similar materials such as $MoSSe$, which has the calculated $\Delta G_H = 1.83\ eV$. A comprehensive benchmark of the calculated hydrogen adsorption free energies for pristine and strained $CrSi_2N_4$ alongside other representative 2D electrocatalysts and bulk Pt is summarized in **Error! Reference source not found.**.

While a pristine basal plane $\Delta G_H$ of 0.46 eV is already competitive as compared to other similar materials and less scarce than the prohibitively expensive bulk Platinum (Pt (111)) catalysts. The true, unmatched power of 2D materials lies in their high susceptibility to physical and chemical modulation. To push the overall catalytic performance from near-ideal to absolute thermoneutral perfection (towards 0.00 eV), researchers used defect introduction and biaxial strain engineering techniques [51][25].

*Strain Engineering*

To study the effect of biaxial strain (in ab-plane) on the electronic band structure and the HER overpotential of the $CrSi_2N_4$ a biaxial strain between +5% and -5% is applied. The structure is allowed to expand along the z-axis to find the minimum energy configuration. The strain was not increased beyond $\pm 5\%$ to avoid the permanent structural deformation. It was noted that the compressive strain increased the interlayer distance. However, the tensile expansive strain slightly reduced the interlayer distance but neither one has posed any significant structural change. This illustrates that the structure can withstand the applied strain. The band structure influenced by applying the biaxial strain is shown in Figure 7. While the transition remains the same (indirect $\Gamma \rightarrow K$), we have observed an increase in PBE bandgap to 1.28 eV by increasing the compressive strain. On the other hand, the, the bandgap decreased upto to 0.25 eV by applying large expansive biaxial strain. Similar results have been reported by Liu et al. for other structures [52]. This establishes the $CrSi_2N_4$ monolayer can be used as a promising, ultra-durable absorber layer for next-generation flexible and wearable tandem photovoltaics and flexible photodetectors. Its structural resilience ensures that device power conversion efficiency and carrier mobility can be reliably maintained under continuous dynamic flexing.

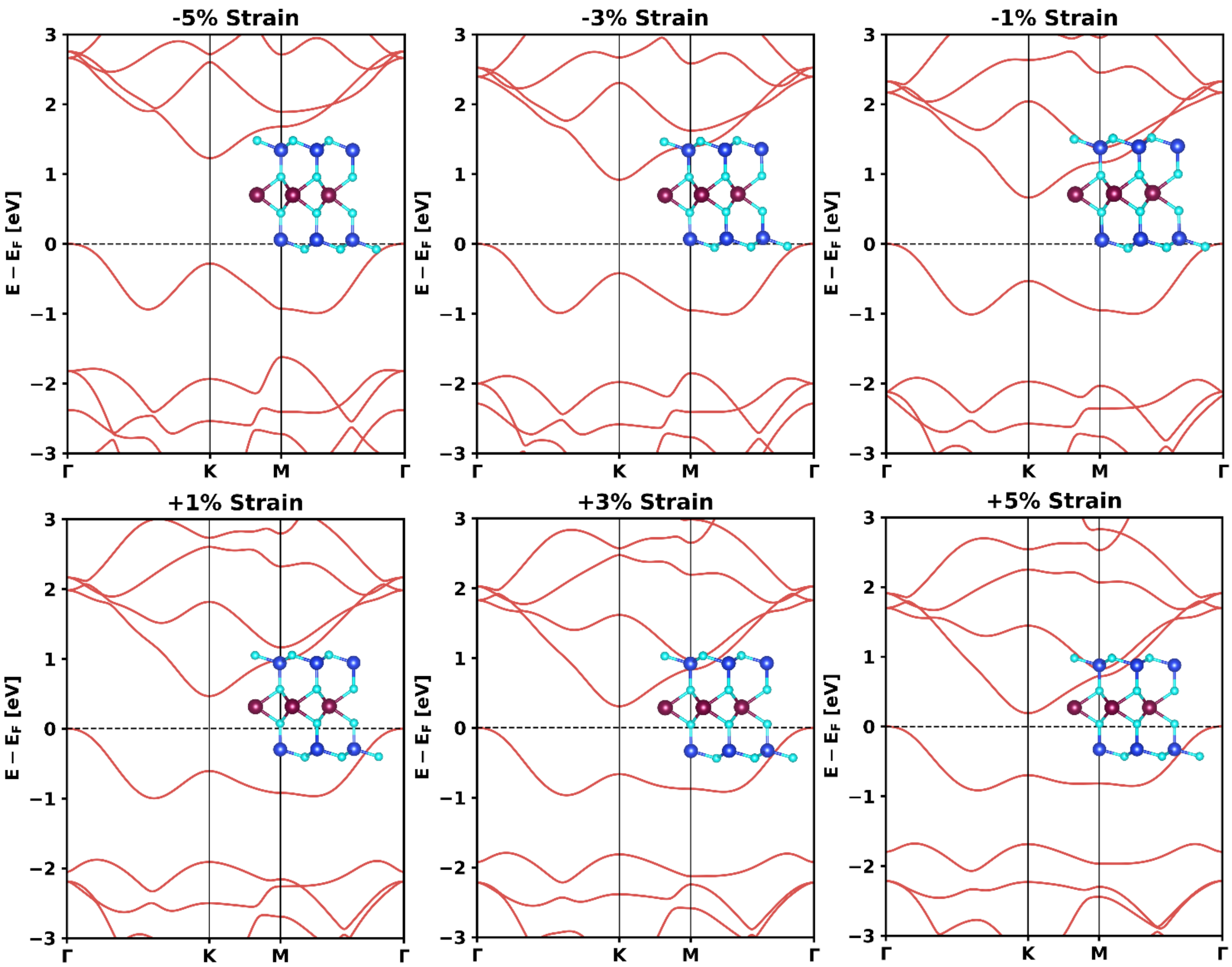

**Figure 7. Strain engineering of $CrSi_2N_4$. Electronic band structure under biaxial strain ranging from -5% (compressive) to +5% (tensile).**

We also calculated the $\Delta G_H$ for hydrogen adsorption at the N-site under different strain conditions. It was observed that applying compressive strain increases the interlayer spacing, which pushes the outer layer farther away from the transition metal atom. This makes the hydrogen evolution reaction (HER) less favorable. In contrast, tensile (expansive) strain improves the reaction by bringing the $\Delta G_H$ value closer to the optimal range. For example, with +5% biaxial strain, $\Delta G_H$ decreases from 1.05 eV to 0.46 eV, pushing its thermodynamic activity significantly closer to the ideal benchmark (as detailed in **Error! Reference source not found.**). Although strain engineering at the N-site does not yet achieve the ideal $\Delta G_H$ value for efficient HER but it clearly demonstrates that the material's properties can be tuned. Table 04 This tunability opens up possibilities for further improvement. Previous studies suggest that additional approaches, such as defect engineering and external electric field tuning, can help bring $\Delta G_H$ closer to the desired value and enhance catalytic performance.

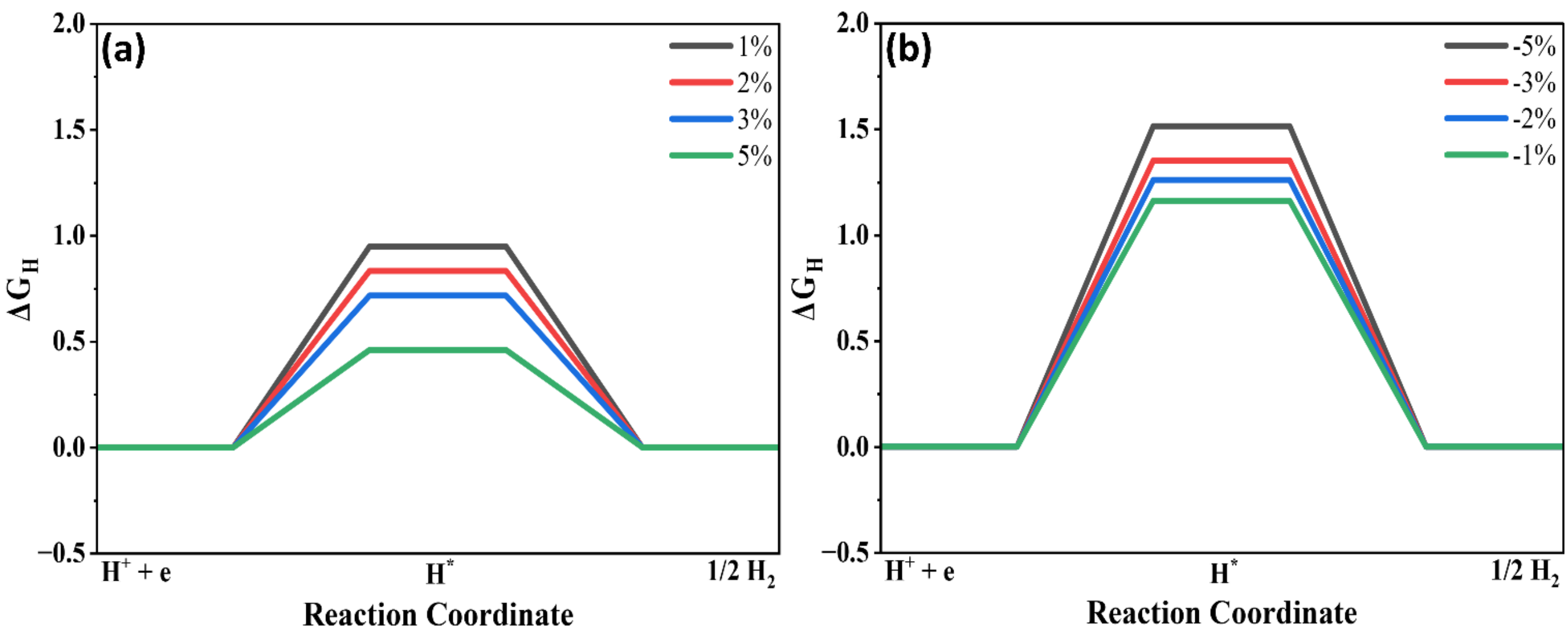

**Figure 8. Modulation of the HER Gibbs free energy ($\Delta G_H$) under applied (a) expansive and (b) compressive biaxial strain, demonstrating enhanced catalytic activity under expansive strain.**

The diverse physical properties calculated for the $CrSi_2N_4$ monolayer highlight its broad potential beyond standalone hydrogen evolution. The high n-type thermoelectric power factor of $CrSi_2N_4$ suggests that it could be integrated into next-generation, low-dimensional waste heat recovery systems or localized cooling for nanoelectronics [53–55]. Simultaneously, its strong optical properties and massive absorption coefficients in both the visible and deep UV ($\sim 2.0 \times 10^6\ cm^{-1}$) region suggest that it is an efficient light-harvesting layer for ultrathin photovoltaic cells and UV photodetectors [16,56,57]. While our strain engineering results successfully demonstrate the tunability of the HER reaction kinetics (reducing $\Delta G_H$ to 0.46 eV under +5% expansive strain), achieving absolute thermoneutrality will require further structural modulation. Future theoretical and experimental investigations should focus on combining strain with targeted surface functionalization, defect engineering (such as the introduction of nitrogen vacancies), or external electric field gating to fully optimize the catalytically active sites. Moreover, combining $CrSi_2N_4$ with other 2D materials such as $MoS_2$ or $WSe_2$ can create van der Waals heterostructures with favourable band alignment, often of Type-II character. These staggered structures promote the effective spatial separation of photo-generated charge carriers, significantly reducing radiative recombination losses [58–60]. Ultimately, the intrinsic stability, wide electronic bandgap, and versatile surface chemistry of $CrSi_2N_4$ establish it as a robust platform for multifunctional energy conversion devices.

**Table 4. Comparison of the calculated hydrogen adsorption Gibbs free energy ($\Delta G_H$) of $CrSi_2N_4$ with other representative electrocatalysts**

| Material | Dimensionality | Adsorption Site | $\Delta G_H$ (eV) | Reference |
|---|---|---|---|---|
| $CrSi_2N_4$ | 2D Monolayer (Pristine) | N-site | 1.05 eV | This work |
| $CrSi_2N_4$ | 2D Monolayer (Strained) | N-site | 0.46 eV | This work |
| $MoSi_2N_4$ | 2D Monolayer (Pristine) | N-site | 2.51 eV | [50] |
| MoSSe | 2D Monolayer (Pristine) | Basal Plane | 1.83 eV | [61] |

| $MoS_2$ | 2D Monolayer | Edge Site | 0.08 eV | [62] |
| --- | --- | --- | --- | --- |
| $MoSi_2N_2P_2$ | 2D Janus Structure (Pristine) | N-site | 1.6 eV | [25] |
| $MoSi_2N_2P_2$ | 2D Janus Structure (Strained) | N-site | 0.7 eV | [25] |
| Pt (111) | Bulk | Surface | ~ 0 eV | Ideal Standard |

**Conclusion**

In summary, a comprehensive first-principles study was conducted to evaluate the structural, electronic, optical, thermoelectric, and electrocatalytic properties of the two-dimensional $CrSi_2N_4$ monolayer. The material exhibits robust dynamic, thermal and mechanical stability at room temperature, underpinned by a high cohesive energy of -8.76 eV/atom, absence imaginary phonon frequencies and no structural deformation in AIMD. Electronic structure calculations confirm a semiconducting nature with an indirect bandgap of 0.58 eV, 2.16 eV with PBE and HSE functionals respectively. This is primarily governed by the covalent hybridization of highly localized Cr-3d and N-2p states near the Fermi level. The material also exhibited remarkable optical performance, including strong absorption in the visible (2.4 eV) and deep-UV (4.0 eV) regions and a high static dielectric constant (15.5), highlighting its potential for optoelectronic and photodetection applications. The thermoelectric transport analysis demonstrated excellent n-type thermoelectric behaviour with a calculated power factor of 3.5 $mWm^{-1}K^{-2}$ at room temperature. This suggests that $CrSi_2N_4$ is a promising candidate for low-dimensional waste heat recovery devices. Furthermore, hydrogen evolution reaction (HER) calculations showed that pristine CrSi2N4 possesses moderate catalytic activity with a hydrogen adsorption free energy of 1.05 eV. Importantly, biaxial tensile strain significantly improved the HER performance, reducing $\Delta G_H$ to 0.46 eV under +5% strain, demonstrating the effectiveness of strain engineering in tuning the catalytic activity. Moreover, the calculated high shear modulus shows its promising integration into the flexible electronic devices without permanent deformation. Overall, the present study establishes $CrSi_2N_4$ as a stable and multifunctional two-dimensional material with promising applications in thermoelectric, optoelectronics and sustainable hydrogen production. These findings provide useful theoretical guidance for future experimental synthesis and device-level investigations of $CrSi_2N_4$ based nanomaterials.

**Acknowledgements**

The authors acknowledge the Ghulam Ishaq Khan Institute of Engineering Sciences and Technology (GIKI) for providing the high-performance computing (HPC) facilities and infrastructure required to perform the extensive computational simulations presented in this study.

**Author contribution**

**Rao Uzair Ahmad**: Conceptualization, Methodology, Software, Formal analysis, Investigation, Data Curation, Writing - Original Draft, Visualization. **Fahd Sikandar Khan:** Validation, Review &

editing. **Nasir Javed:** Supervision, Conceptualization, Validation, Resources, Review & Editing, Project administration.

## Declaration of Competing Interest

The authors declare that they have no known competing financial interests or personal relationships that could have appeared to influence the work reported in this paper.

## Data Availability

The raw and processed computational data generated during this study (including optimized structural coordinates and output files) will be made available by the corresponding author upon reasonable request.